\documentclass[aps,twocolumn,noshowkeys,superscriptaddress,noshowpacs]{revtex4}

\usepackage{dcolumn}
\usepackage{amsmath}
\usepackage{bm}
\usepackage{nicefrac}
\usepackage{graphicx}
\usepackage{siunitx}
\usepackage{epsfig}
\usepackage{epstopdf}
\usepackage{xcolor}

\newcommand{\spinel}{MgAl$_\text{2}$O$_\text{4}$} 
\newcommand{\nife}{NiFe$_\text{2}$O$_\text{4}$}
\newcommand{\Ar}{Ar\,:\,O$_2$ ratio~}

\begin{document}

\title{Influence of structure and cation distribution on magnetic anisotropy and damping in Zn/Al doped nickel ferrites}
\date{\today}

\author{Julia Lumetzberger}
\email{julia.lumetzberger@jku.at; Phone: +43-732-2468-9651; FAX: -9696}
\affiliation{Johannes Kepler University Linz, Institute for Semiconductor and Solid State Physics, Altenberger Strasse 69, 4040 Linz, Austria}
\author{Martin Buchner}
\affiliation{Johannes Kepler University Linz, Institute for Semiconductor and Solid State Physics, Altenberger Strasse 69, 4040 Linz, Austria}
\author{Santa Pile}
\affiliation{Johannes Kepler University Linz, Institute for Semiconductor and Solid State Physics, Altenberger Strasse 69, 4040 Linz, Austria}
\author{Verena Ney}
\affiliation{Johannes Kepler University Linz, Institute for Semiconductor and Solid State Physics, Altenberger Strasse 69, 4040 Linz, Austria}
\author{Wolfgang Gaderbauer}
\affiliation{Christian Doppler Laboratory for Nanoscale Phase Transformations, Johannes Kepler University Linz, Center for Surface and Nanoanalytics, Altenberger Strasse 69, 4040 Linz, Austria}
\author{Ni\'{e}li Daff\'{e}}
\affiliation{Swiss Light Source (SLS), Paul Scherrer Institut, 5232 Villigen PSI, Switzerland}
\author{Marcos V. Moro}
\affiliation{Department of Physics and Astronomy, \si{\angstrom}ngstr{\"o}m Laboratory, Uppsala University, Box 516, SE-751 20 Uppsala, Sweden}
\author{Daniel Primetzhofer}
\affiliation{Department of Physics and Astronomy, \si{\angstrom}ngstr{\"o}m Laboratory, Uppsala University, Box 516, SE-751 20 Uppsala, Sweden}
\author{Kilian Lenz}
\affiliation{Helmholtz-Zentrum Dresden -- Rossendorf, Institute of Ion Beam Physics and Materials Research, Bautzner Landstr. 400, 01328 Dresden, Germany}
\author{Andreas Ney}  
\affiliation{Johannes Kepler University Linz, Institute for Semiconductor and Solid State Physics, Altenberger Strasse 69, 4040 Linz, Austria}

\begin{abstract}
An in-depth analysis of Zn/Al doped nickel ferrites grown by reactive magnetron sputtering is relevant due to their promising characteristics for applications in spintronics. The material is insulating and ferromagnetic at room temperature with an additional low magnetic damping. By studying the complex interplay between strain and cation distribution their impact on the magnetic properties, i.e. anisotropy, damping and g-factor is unravelled. In particular, a strong influence of the lattice site occupation of Ni$^{2+}_{\text{Td}}$ and cation coordination of Fe$^{2+}_{\text{Oh}}$ on the intrinsic damping is found. Furthermore, the critical role of the incorporation of Zn$^{2+}$ and Al$^{3+}$ is evidenced by comparison with a sample of altered composition. Especially, the dopant Zn$^{2+}$ is evidenced as a tuning factor for Ni$^{2+}_{\text{Td}}$ and therefore unquenched orbital moments directly controlling the g-factor. A strain-independent reduction of the magnetic anisotropy and damping by adapting the cation distribution is demonstrated.
\end{abstract}

\maketitle

\section*{I. Introduction}
In spintronics one aims to obtain pure spin currents as an additional degree of freedom in logic circuits aside from electric charges \cite{WA01}. By spin pumping \cite{TBB02} it is feasible to induce a spin current from ferromagnetic materials into adjacent non-magnetic layers. To ensure pure spin currents and exclude charge currents ferromagnetic insulators are the materials of choice. However, ferromagnetic insulators with low intrinsic magnetic damping are sparse \cite{HE15}. The most commonly used material is yttrium iron garnet (YIG) \cite{OKK14, CLZ14}. YIG has two major drawbacks, i.e. the complex garnet structure, which is almost exclusively grown on gadolinium gallium garnet (GGG) substrates and its weak magnetoelastic response. It could be replaced by cubic Zn/Al doped nickel ferrite (NiZAF) thin films grown on \spinel, which were reported to have comparably favorable magnetic properties \cite{E17} as YIG: ferromagnetic at room temperature and a low intrinsic damping. Furthermore, the static magnetic analysis shows a soft magnetic behavior with a low coercive field, which makes the material relevant for high frequency applications. The combination of these properties makes it especially suitable for applications in magnetoelectric and acoustic spintronics.  \\
NiZAF is based on nickel-ferrite \nife, which ideally grows in the cubic inverse spinel crystal structure, i.e. Ni occupies octahedral (Ni$_{\text{Oh}}$) lattice sites and Fe is located in a $1$\,:\,$1$ ratio either at tetrahedral (Fe$_{\text{Td}}$) or octahedral (Fe$_{\text{Oh}}$) lattice sites. This is explained in more detail in section VI [see Fig.\,\ref{figure5}(a)].
Thin film growth of NiZAF on a MgAl$_2$O$_4$ (001) substrate with a normal spinel crystal structure gives rise to disorder between the tetrahedral A (Ni$^{2+}$) and octahedral B (Fe$^{3+}$) sites. Therefore Ni and Fe with a valency of 2+ (Ni$^{2+}_{\text{Td}}$ and Fe$^{2+}_{\text{Oh}}$) are introduced into the lattice. Previous work on the single ion model of ferrite magnetism \cite{D11} shows a negative impact of tetrahedrally coordinated Ni$^{2+}_{\text{Td}}$ on the magnetic damping due to unquenched orbital moments. This effect has not been found for octahedrally coordinated Ni$^{2+}_{\text{Oh}}$. Additionally, Fe$^{2+}_{\text{Oh}}$ increases the magnetic damping due to hopping. Thin film growth of co-sputtered \nife\,on the same substrate confirms the presence of Fe$^{2+}_{\text{Oh}}$ in this system \cite{K14}.\\
The structural and magnetic properties can be controlled by doping with Zn and Al, which predominantly substitute Ni and Fe, respectively. Since Zn$^{2+}$ prefers the tetrahedral coordination it prevents Ni$^{2+}$ from occupying these sites and therefore, predominantly Ni$^{2+}_{\text{Oh}}$ is formed \cite{N67}. Growth of highly strained \nife \, thin films \cite{K14} showed an increase in damping caused by defects. Thus, to reduce the strain, smaller Al$^{3+}$ cations are added to substitute for Fe$^{3+}$. A doping percentage of 0.8 was reported to be the best compromise between magnetostriction and saturation magnetization for bulk polycrystalline NiZAF \cite{L13}. \\
Substituting ferromagnetic with non-ferromagnetic elements in NiZAF results in a reduced saturation magnetization $M_{\text{S}} \sim$\,\SI{110}{\kilo\ampere/\meter} and Curie temperature $\sim$\,\SI{450}{\kelvin} \cite{E17} compared to undoped \nife \, with $M_{\text{S}}$ up to $300$\,\si{\kilo\ampere/\meter} and a significantly higher Curie temperature of $T_{\text{C}} = 860$\,\si{\kelvin} \cite{H12}. In turn, NiZAF exhibits a smaller ferromagnetic resonance (FMR) linewidth down to $\sim$\,\SI{0.82}{\milli\tesla} resulting in a Gilbert damping parameter of $\alpha \sim 3 \times 10^{-3}$ \cite{E17} in comparison to a linewidth $\mu_0 \Delta H_{\text{pp}} = 35$\,\si{\milli\tesla} of \nife \cite{H12}. Furthermore, in contrast to a coercivity of $H_{\text{c}} = 100$\,\si{\milli\tesla} \cite{K14} for undoped \nife \, a reduced coercive field of $\sim$\,\SI{0.2}{\milli\tesla} \cite{E17} has been reported. Zn/Al doping of \nife \, is, therefore, an advantageous trade-off since low damping and coercivity are desired for applications in spintronics. \\
NiZAF was studied for the correlation between composition, strain, cation distribution, magnetic anisotropy and intrinsic damping. This manuscript is structured as follows: in section II the experimental details of the sample fabrication are provided together with the experimental details of the various techniques used throughout this work. At first the structural characterization is discussed in section III followed by the static and dynamic magnetic analysis in section IV. In section V the cation distribution in the material is studied in detail by comparing the experimental spectra with multiplet ligand field simulations. A comparison to a sample with intentionally altered stoichiometry is given in section VI to highlight the governing mechanism in the material system. Finally, a short conclusion is given in section VII.

\section*{II. Experimental Details}
In this work NiZAF was fabricated using reactive magnetron sputtering (RMS) with a nominal target composition of Ni$_\text{0.65}$Zn$_\text{0.35}$Al$_\text{0.8}$Fe$_\text{1.2}$O$_\text{4}$ as suggested in Ref.\cite{E17}, in which pulsed laser deposition (PLD) was used. The epitaxial thin films were grown in an ultra high vacuum (UHV) chamber with a base pressure of $2 \times 10^{-9}$\,\si{\milli\bar}. A doubleside polished single crystalline spinel [\spinel(001)] was chosen as a substrate. The data of the growth optimization including temperature, \Ar and thickness variation and annealing can be found in the data repository \cite{LBN19-S}. The best results were achieved with the highest possible heater temperature corresponding to a sample temperature of $\sim$\,\SI{525}{\degreeCelsius}, an \Ar of ($10$\,:\,$0$)\,\si{sccm} to ($10$\,:\,$0.1$)\,\si{sccm}, a magnetron power of \SI{30}{\watt} and a working pressure of $4 \times 10^{-3}$\,\si{\milli\bar}. Additionally, a sample with an adjusted composition, i.e., increased Zn while maintaining the Al concentration was fabricated. During that growth a sample temperature up to $\sim$\,\SI{600}{\degreeCelsius} could be achieved. \\
The structural characterization of the samples was done by X-ray diffraction (XRD) measurements with a \textit{Pananalytical X'Pert MRD} and a \textit{Seifert XRD3003} recording $\omega - 2 \theta$ scans and symmetric as well as asymmetric reciprocal space maps (RSM). The surface roughness was characterized with an atomic force microscope (AFM). The interface on an atomic scale was checked by transmission electron microscopy (TEM) using a \textit{Jeol JEM-2200 FS}. Furthermore, the chemical composition was obtained with energy dispersive X-ray spectroscopy (EDX) in the TEM system. Additionally, the composition was determined more accurately by ion beam analysis, i.e. Rutherford backscattering spectrometry (RBS) using \SI{2}{\mega\electronvolt} He$^{+}$ and \SI{10}{\mega\electronvolt} $^{12}$C$^{3+}$ primary beams at the Tandem Laboratory at Uppsala University. To disentangle the element specific contributions, the spectra were analyzed using the SIMNRA software \cite{M97}. Details of the experimental setup are described elsewhere \cite{M19}.
The static magnetic properties were measured with integral superconducting quantum interference device (SQUID) magnetometry by a \textit{Quantum Design MPMS-XL5} system applying the magnetic field in the film plane (IP) as well as out-of-plane (OOP). The magnetic behavior in a field range of $\pm 5$\,\si{\tesla} from room temperature (RT) down to \SI{2}{\kelvin} and the temperature dependence of the magnetization up to \SI{400}{\kelvin} were measured. The data were background corrected for the contribution of the diamagnetic substrate and known artefacts were carefully avoided \cite{SSA11, B18}.\\
The dynamic magnetic properties were analyzed with a broadband vector network analyzer ferromagnetic resonance (VNA-FMR) setup at room temperature at the HZDR, researching the polar and azimuthal angular dependence of resonance position and linewidth. Furthermore, the frequency dependence was measured to verify the g-factor and disentangle the different contributions to the magnetic damping. The angular dependence was cross-checked with a conventional X-band FMR setup and the data can be found in \cite{LBN19}. \\
Finally, X-ray absorption (XAS) and X-ray magnetic circular dichroism (XMCD) spectra at the Ni and Fe L$_{3,2}$ edges were measured at the X-Treme beamline at the Swiss Light Source (SLS) \cite{PFS12}. The XMCD spectra were obtained in total electron yield (TEY) with \SI{20}{\degree} grazing incidence at RT and a field of \SI{6.8}{\tesla}. 
The XMCD(H) curves were measured in a field range of $\pm$\,\SI{6.8}{\tesla} in total fluorescence yield (TFY) at normal incidence, RT and at the maximum of the XMCD peaks sensitive to the site occupancy of the cation in the crystallographic structure. 
The raw data of the measurements can be found in the data repository \cite{LBN19}.

\section*{III. Structural Properties}

\begin{figure}[h]
	\centering
		\includegraphics[width=0.41\textwidth]{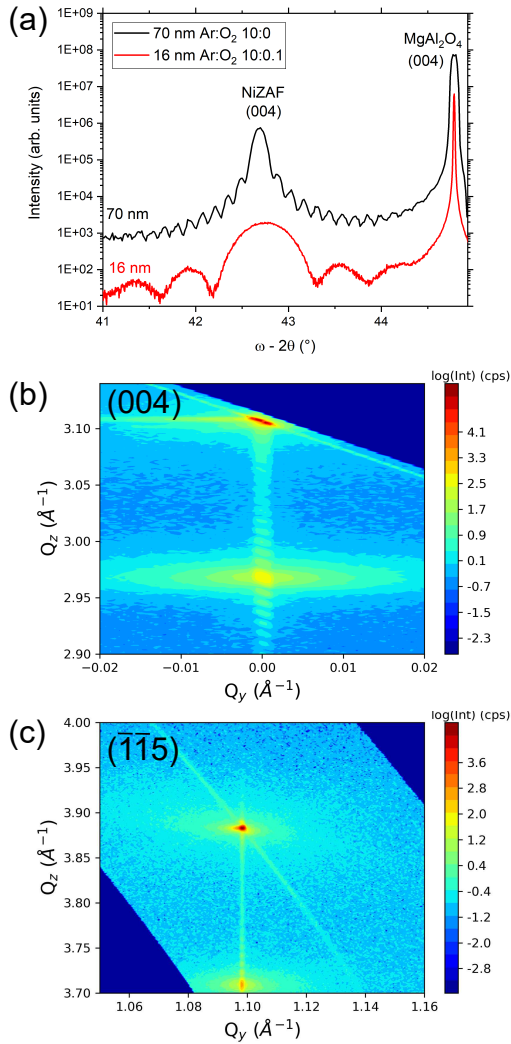}
		\vspace{-0.2cm}
	\caption{Structural properties of NiZAF measured with XRD on a \SI{70}{\nano\meter} sample. (a) shows a symmetric $\omega-2 \theta$ scan in comparison to a \SI{16}{\nano\meter} sample, (b) symmetric RSM along ($004$) and (c) asymmetric RSM along ($\bar{1}\bar{1}5$).}
	\label{figure1}
\end{figure} 

The crystalline structure of the material is routinely analyzed by symmetric $\omega-2 \theta$ scans using XRD. Results for two samples with different thicknesses are shown in Fig.\,\ref{figure1}(a). Both samples have the (004) reflection at $42.67$\,\si{\degree} with respect to the substrate and show Laue oscillations. The film reflection with a small full width at half maximum (FWHM) around $0.13$\,\si{\degree} for the \SI{70}{\nano\meter} sample corresponds to a perpendicular lattice parameter of $a_{\perp} = (8.47 \pm 0.01)$\,\si{\angstrom}. The symmetric RSM along the ($004$) plane and the asymmetric RSM along the ($\bar{1}\bar{1}5$) plane are shown in Figs.\,\ref{figure1}(b) and \ref{figure1}(c), respectively. The reflection from the \spinel \, substrate corresponds to a lattice parameter of $a_{\text{sub}} = 8.08$\,\si{\angstrom}, which has a lattice mismatch of $\sim$\,\SI{2}{\percent} to bulk NiZAF ($a_0 = 8.24$\,\si{\angstrom}) \cite{L13}. The asymmetric scan reveals an in-plane lattice parameter of $a_{\parallel} = (8.09 \pm 0.01)$\,\si{\angstrom}, i.e. fully strained with respect to the substrate. A calculation of the tetragonal distortion yields $c/a = 1.047 \pm 0.003$ confirming a clearly strained material. Additionally, the unit cell volume of the strained material is $a_{\parallel}^2 \cdot a_{\perp} = (554 \pm 2)$\,\si{\cubic\angstrom}, which is equal to a reduction of the unit cell volume by only $\sim$\,\SI{1}{\percent} with respect to the bulk $a_0^3 =$\,\SI{559.5}{\cubic\angstrom}. Even though the samples are significantly strained, they maintain an excellent crystal quality evidenced by the well pronounced Laue oscillations visible in the line scan as well as the RSM indicating a fully strained homogeneous crystal growth for a thickness of up to \SI{70}{\nano\meter}.

\begin{figure*}[ht]
	\centering
	\includegraphics[width=0.84\textwidth]{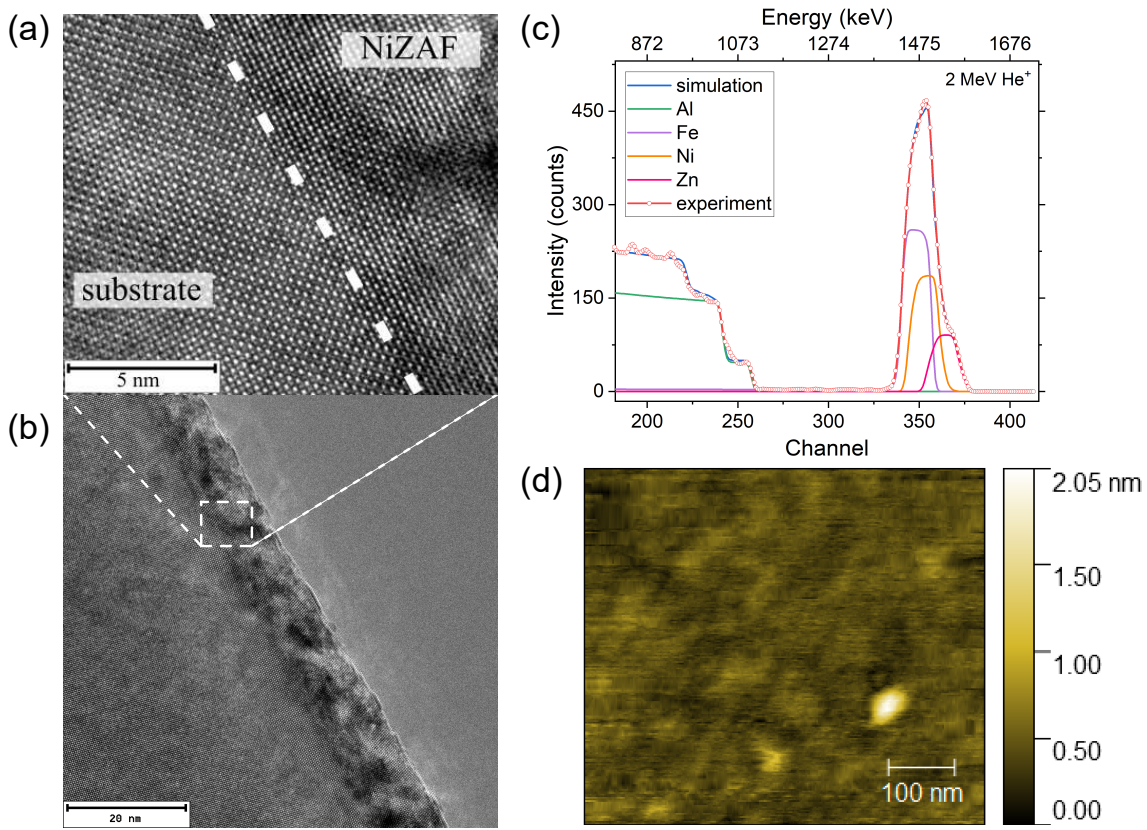}
	\vspace{-0.2cm}
	\caption{Surface and interface properties of a \SI{70}{\nano\meter} NiZAF sample. (a) a high-resolution cross section TEM image showing the interface of film and substrate on an atomic scale, (b) the film on a larger scale. The RBS spectrum from a \SI{2}{\mega\electronvolt} He$^+$ beam to determine the chemical composition analysed by SIMNRA is depicted in (c). (d) shows an AFM scan of the surface.}
	\label{figure2}
\end{figure*} 

In Fig.\,\ref{figure2}(a) the high resolution cross-section TEM image along the $\left\langle 100\right\rangle$ direction shows a sharp interface transition between film and substrate on an atomic scale. No indication for dislocations or defects can be found in the measurements on various length scales, further proving the excellent structural quality of the coherently strained material. A clear separation between film and substrate is visible at the sample interface on a larger scale, depicted in Fig.\,\ref{figure2}(b). However, the film exhibits a clearly enhanced strain contrast, which corroborates the findings of the RSM [see Fig.\,\ref{figure1}(b)]. Additional annealing had no influence on the structural quality nor the magnetic properties further supporting a low-defect material, in contrast to previous works on epitaxially grown spinel ferrite thin films \cite{H12,S01}.\\
The analysis of the chemical composition with EDX in the TEM system (not shown, see repository \cite{LBN19}) revealed a ratio of ($\text{Fe}:\text{Ni}:\text{Zn}$)\,($1:0.55:0.22$). In comparison to the nominal values of ($\text{Fe}:\text{Ni}:\text{Zn}:\text{Al}$)\,($1:0.54:0.29:0.66$) (all values are normalized with respect to the Fe amount in the material) a clear reduction of Zn content is observed. The deficiency in Zn can be explained by its high volatility, which is even more enhanced by the high growth temperatures \cite{ASY16}. The aluminium content cannot be reliably determined by EDX due to the difference in atomic number in comparison to the other metals. Therefore, the elemental composition was refined by an in-depth ion beam analysis using RBS with \SI{2}{\mega\electronvolt} He$^+$ primary particles, which makes an accurate determination of the concentration depth profile of lighter elements like Al feasible. From these measurements a chemical composition of ($\text{Fe}:\text{Ni}:\text{Zn}:\text{Al}$)\,($1:0.62:0.26:0.71$) is determined [see Fig.\ref{figure2}(c)]. The Zn deficiency is less than suggested from EDX, however the amount of Ni is increased compared to the nominal values from the composite target. The Al content is slightly increased, indicating only a small deviation from the nominal values. These findings suggest that the stoichiometric transfer for this composition by RMS is comparable to PLD \cite{E17} since both results are in reasonable agreement with the target composition. Note that the Al content cannot be correlated due to the lack of data from the PLD samples in Ref.\,\cite{E17}. Furthermore, by analyzing the sample with a \SI{10}{\mega\electronvolt} $^{12}$C$^{3+}$ ion beam, a more reliable result for the heavier elements of ($\text{Fe}:\text{Ni}:\text{Zn}:\text{Al}$)\,($1:0.6:0.23:0.83$) could be obtained due to the better mass separation. Thus confirming previous results, i.e. a good incorporation of Ni with a slight deficiency in Zn. The increased Al content is confirmed from this measurement. From the spectrum in Fig.\,\ref{figure2}(c) no drastic gradient is visible in the element concentration. This behavior is supported for Ni, Zn and Fe by the C ion beam \cite{LBN19}.\\
In Fig.\,\ref{figure2}(d) an AFM image of the surface of a $550 \times 550$\,\si{\nano\meter\squared} area is shown. A flat surface with neither cracks nor holes and a low average maximum roughness peak height $R_{pm} = 40$\,\si{\pico\meter} as well as a root mean square roughness of $R_{q} = 24$\,\si{\pico\meter} is found. This is further supported by the maximum height deviation of \SI{2}{\nano\meter} within the image, which corresponds to only 2 monolayers of NiZAF. The smooth surface is ideal for growing heterostructures making NiZAF a perfect choice for spin pumping devices with an insulating ferromagnetic material.

\begin{figure}[h]
	\centering
		\includegraphics[width=0.41\textwidth]{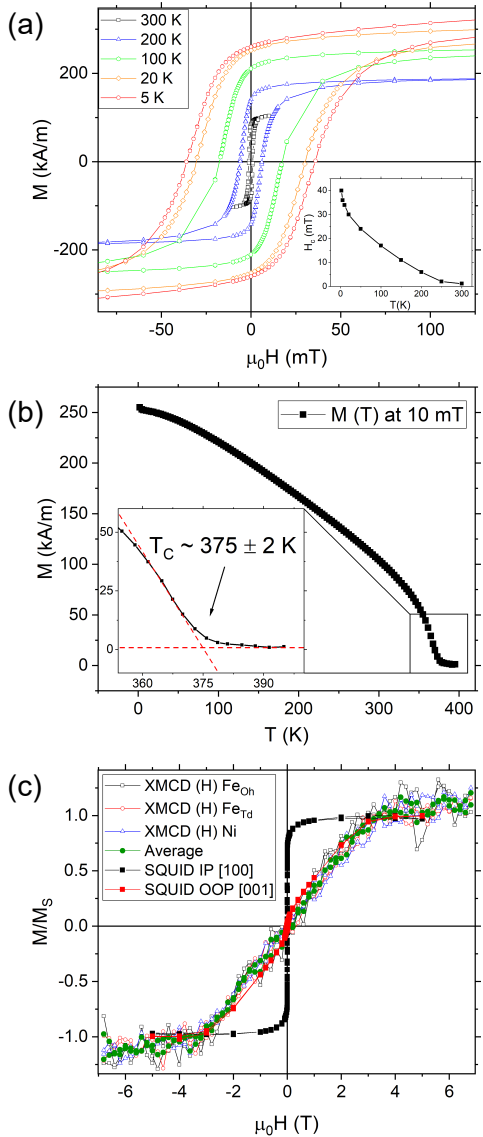}
		\vspace{-0.2cm}
	\caption{Static magnetic properties obtained from integral SQUID magnetometry. (a) shows the hystereses at exemplary temperatures between \SI{300}{\kelvin} and \SI{2}{\kelvin} with an inset giving a detailed look at the coercive field in this temperature range. (b) depicts the temperature dependence of the magnetization from \SIrange{2}{400}{\kelvin} at \SI{10}{\milli\tesla}. An estimation of the Curie temperature is given in the inset. In (c) the XMCD (H) of Fe$_\text{Oh}$, Fe$_\text{Td}$, Ni and an average of all three in comparison with the integral IP $\left\langle 100\right\rangle$ and OOP $\left\langle 001\right\rangle$ SQUID measurements are plotted.}
	\label{figure3}
\end{figure} 

\section*{IV. Magnetic Properties}

The static and dynamic magnetic behavior of the material is measured by SQUID magnetometry and FMR, respectively. In Fig.\,\ref{figure3}(a) the magnetic hystereses at selected temperatures are shown. The inset depicts the dependence of the coercive field for all investigated temperatures between \SI{300}{\kelvin} and \SI{2}{\kelvin}. A small coercive field of $H_{\text{c}} = 1.2 $\,\si{\milli\tesla} at \SI{300}{\kelvin} is determined, which is further evidence for a material devoid of pinning centers. A steady increase from $H_{\text{c}} = 1.2 $\,\si{\milli\tesla} at \SI{300}{\kelvin} up to $H_{\text{c}} = 40 $\,\si{\milli\tesla} at \SI{2}{\kelvin} was determined, as can be seen in the inset of Fig.\,\ref{figure3}(a). For low temperatures the typical increase of remanence and coercivity is observed. \\
The volume of the film is taken into account for the magnetization values by thickness determination from the Laue oscillations [see Fig.\,\ref{figure1}(a)] and the area of the used sample piece. Accordingly, a saturation magnetization of $M_{\text{S}} = (118 \pm 12)$\,\si{\kilo\ampere/\meter} is obtained from the SQUID data at RT. This value matches the bulk value of $\sim$\,\SI{120}{\kilo\ampere/\meter} \cite{L13} and previous values for thin films $\sim$\,\SI{120}{\kilo\ampere/\meter} \cite{E17}. The Curie temperature of $T_{\text{C}} = (375 \pm 2)$\,\si{\kelvin} was estimated from the $M(T)$ curve at \SI{10}{\milli\tesla} shown in the inset of Fig.\,\ref{figure3}(b). Compared to PLD samples with $T_{\text{C}} = $\,\SI{450}{\kelvin} \cite{E17} the Curie temperature is slightly reduced. Nevertheless, it is well above \SI{300}{\kelvin} making ferromagnetic applications at RT feasible. \\
In addition, the directional dependence of the static magnetic properties was studied. In Fig.\,\ref{figure3}(c) the difference in saturation field between IP $\left\langle 100\right\rangle$ and OOP $\left\langle 001\right\rangle$ $M(H)$ curves measured by SQUID is shown in comparison with XMCD (H) curves of  Fe$_\text{Oh}$, Fe$_\text{Td}$, Ni and an average of all three. For now, only the comparison of the different contributions to the average with respect to the SQUID data is of importance and the XMCD spectra will be discussed in more detail further below. A large uniaxial OOP anisotropy field of $\mu_0 H_{\text{k}} > 3$\,\si{\tesla} is obtained, which cannot be solely explained by shape anisotropy, which is $\mu_0 M_{\text{S}} = $\,\SI{0.15}{\tesla}. Furthermore, the high OOP anisotropy field of more than \SI{3}{\tesla} is evident in the XMCD(H) curves. The average of the element selective hystereses shows the same behavior as the OOP $M(H)$ curve of the integral SQUID measurements. Fully strained \nife \ with lower crystal quality, on the other hand shows no contribution to the OOP anisotropy field except for shape \cite{K14}. The large $\mu_0 H_{\text{k}}$ presumably originates from the magnetocrystalline anisotropy of the highly strained inverse spinel crystal structure. This is supported by previous work on less-strained Zn/Al doped nickel ferrite ($c/a = 1.035$) with a smaller out-of-plane anisotropy ($\mu_0 H_{\text{k}} > 1$\,\si{\tesla}) in Ref.\,\cite{E17}. At first sight the major change in anisotropy from $\mu_0 H_{\text{k}} > 1$\,\si{\tesla} \cite{E17} to $\mu_0 H_{\text{k}} > 3$\,\si{\tesla} seems to be caused by a change in the $c$ lattice parameter from $a_{\perp} = 8.36$\,\si{\angstrom} \cite{E17} to $a_{\perp} = 8.47$\,\si{\angstrom} for the present film. The fact that, in the present sample the Al content is slightly higher than nominal together with the lack of knowledge about the actual Al content in Ref.\,\cite{E17} indicates that another factor could play a decisive role for the increased damping.

\begin{figure}[h]
	\centering
	\includegraphics[width=0.41\textwidth]{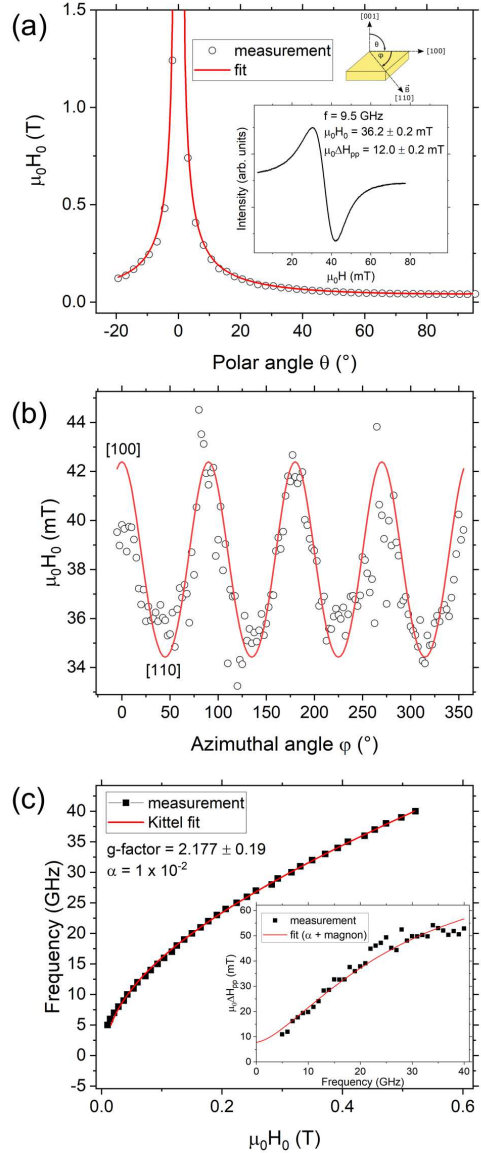}
	\vspace{-0.2cm}
	\caption{Dynamic magnetic properties including (a) the polar angular dependence from \SIrange{-20}{90}{\degree} and (b) the azimuthal angular dependence from \SIrange{0}{360}{\degree} with an anisotropy fit (red line) for the resonance position in both dependences. The inset in (a) depicts the smallest FMR line measured with a conventional setup at $f= 9.5$\,\si{\giga\hertz} in the easy axis. (c) shows the frequency dependence of the resonance position as well as linewidth (inset) in a range from \SIrange{5}{40}{\giga\hertz}.}
	\label{figure4}
\end{figure}

Angle- and frequency-dependent FMR measurements have been performed at RT to obtain accurate values for magnetic anisotropy and damping. The FMR spectra were fitted by Lorentzians to obtain the resonance field ($\mu_0H$) and peak-to-peak linewidth ($\mu_0\Delta H_\mathrm{pp}$). Figure \ref{figure4}(a) shows the polar angular dependence measured at $f=9.5$\,GHz by VNA-FMR, where $\theta_H=0^\circ$ denotes the out-of-plane direction, i.e.\ the hard axis [see Fig.\,\ref{figure4}(a)]. The resonance field in this direction was larger than the available field range of the electromagnet. Hence, no resonance could be found around the hard axis. The inset shows a typical FMR spectrum taken for field in-plane, i.e., $\theta_H=90^\circ$, using a conventional X-band resonator with field-modulation technique. The fit reveals a resonance field of $\mu_0H=(36.2\pm 0.2)$\,\si{\milli\tesla} and a peak-to-peak linewidth of $\mu_0\Delta H_\mathrm{pp} = (12.0 \pm 0.2)$\,\si{\milli\tesla}. Compared to that the linewidth of the bulk material is four times higher, i.e., $\mu_0\Delta H_\mathrm{pp} = 43$\,\si{\milli\tesla} \cite{L13}.
Figure \ref{figure4}(b) shows the corresponding azimuthal angular dependence clearly indicating a cubic (fourfold) magnetocrystalline anisotropy with easy axes along the $\left\langle 110\right\rangle$ directions. A comparison with bulk shows that the IP easy axis switches from $\left\langle 111\right\rangle$ to $\left\langle 110\right\rangle$ directions for thin film growth in agreement with \cite{E17}. \\
To fit the angular dependences in order to determine the cubic IP ($K_{4||}/M_\mathrm{s}$) and OOP ($K_{4\perp}/M_\mathrm{s}$) anisotropy fields as well as the intrinsic uniaxial OOP ($K_{2\perp}/M_\mathrm{s}$), IP ($K_{2||}/M_\mathrm{s}$) and shape contributions, the following free energy density function was used \cite{F98}:
\begin{align}
	 	F  = & - M_\mathrm{s} H_{0}[\sin\theta\sin\theta_H\cos(\phi - \phi_H) + \cos\theta \cos\theta_H] \nonumber \\
	& - (2\pi M_\mathrm{s}^2 - K_{2\perp} ) \sin^2(\theta) \nonumber \\
	& - K_{2\parallel} \sin^2(\theta) \cos^2(\phi) \nonumber \\
	& - \frac{K_{4\perp}}{2} \cos^4(\theta) \nonumber \\
	& - \frac{K_{4\parallel}}{8} [3 + \cos(4\phi)]\sin^4\theta  \quad.
\end{align}
Here $\theta, \theta_H$ and $\phi, \phi_H$ denote the OOP and IP angles of the magnetization and external magnetic field, respectively. 
To account for a possible deviation from the cubic system due to the high strain in the material the free energy density of the lower-symmetric tetragonal crystal structure was used. 
The fits of the experimental polar and azimuthal angular dependences to the resonance condition [red curves in Figs.\,\ref{figure4}(a) and \ref{figure4}(b)] result in the following anisotropy fields: The effective magnetization is $\mu_0 M_\mathrm{eff} = \mu_0 M_\mathrm{s} - \frac{2K_{2\perp}}{M_\mathrm{s}}= 2.5$\,\si{\tesla}. Using $M_\mathrm{s}$ determined by SQUID the uniaxial OOP anisotropy field $2K_{2\perp}/M_\mathrm{s}$ can be calculated as \SI{2.35}{\tesla}, which correlates well with the anisotropy field estimated from SQUID [see Fig.\,\ref{figure4}(b)]. 
This is in agreement with the expectation that a strong hard axis in a material with a low saturation magnetization has to stem from a strain induced perpendicular uniaxial anisotropy component. The uniaxial IP anisotropy field $K_{2\parallel}/M_\mathrm{s}$ is disregarded, since the visible scatter of the data points makes potential contribution not significant. Furthermore, the fits reveal that $K_{4\parallel}$ and $K_{4\perp}$ are the same. Hence, the uniform cubic anisotropy field is $2K_4/M_\mathrm{s} = - 4$\,\si{\milli\tesla}. Previous work showed an even stronger preference for the $\left\langle 110\right\rangle$ easy axis with an in-plane anisotropy of $\sim 10$\,\si{\milli\tesla} for a less-strained material ($c/a = 1.035$) grown by PLD \cite{E17}.
The $g$-factor of $g=2.18 \pm 0.19$ was obtained from the frequency-dependence of the resonance field depicted in Fig.\,\ref{figure4}(c) by using the Kittel equation \cite{Z07}. This infers to a lower contribution of the orbital momentum in comparison to Ref.\,\cite{E17}, where $g=2.29 \pm 0.09$ was reported.
The inset of Fig.\,\ref{figure4}(c) depicts the frequency dependence of the linewidth. A monotonic non-linear dependence is visible, which is a clear sign for a two-magnon scattering (TMS) contribution to the damping in addition to the intrinsic Gilbert damping \cite{L03, L06}, which is linear in frequency. The fit (red line) includes both components resulting in a Gilbert damping parameter of $\alpha = 1 \times 10^{-2}$ and a TMS factor of $\Gamma_{100}=0.04\times 10^{-8}$~s$^{-1}$, respectively. Even neglecting the TMS contribution, the pure Gilbert damping is still higher by an order of magnitude than previously reported \cite{E17}. Additionally, an inhomogeneous contribution of $B_{\text{inhom}} = $\,\SI{7.8}{\milli\tesla} is apparent from the linewidth dependence. Since the material and substrate are both insulating and without additional cap layers, broadening by eddy current damping and/or spin pumping can be excluded.\\
In a first evaluation of the magnetic properties in comparison to the structural analysis the following observation is made: The higher strained RMS material ($c/a = 1.047 \pm 0.001$) has a lower cubic anisotropy $2K_4/M_\mathrm{s}$ of $4$\,\si{\milli\tesla}, yet shows a significant increase in magnetocrystalline anisotropy evidenced by the high $2K_{2\perp}/M_\mathrm{s}$ of \SI{2.35}{\tesla}. In turn, this leads to a broadening of the linewidth and increased damping with a non negligible contribution of TMS at higher frequencies. Even though, the structural analysis shows an excellent crystal structure with a low amount of defects at least on a local scale, a significant inhomogeneous component is detectable by sensitive magnetic measurements.

\section*{V. Cation distribution}

\begin{figure*}[ht]
	\centering
	\includegraphics[width=0.84\textwidth]{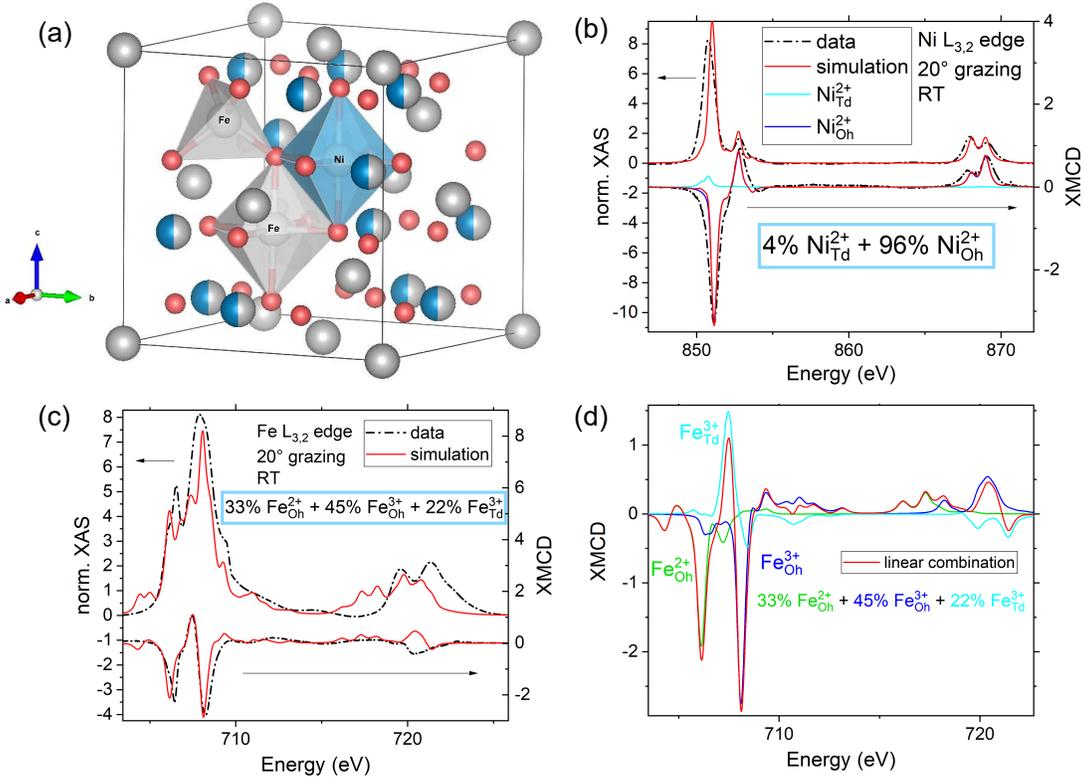}
	\vspace{-0.2cm}
	\caption{(a) Sketch of the inverse spinel crystal structure of \nife \, as the underlying structure for NiZAF. Total electron yield XAS and XMCD spectra in grazing incidence of \SI{20}{\degree} at RT and simulations done with CTM4XAS \cite{SG10} are depicted at (b) the Ni L$_{3,2}$ and (c) the Fe L$_{3,2}$ edges. (d) shows the cation distributions of the simulated weighted linear combination of the XMCD at the Fe L$_{3,2}$ edges.}
	\label{figure5}
\end{figure*} 

To verify if strain is causing the increased damping, the cation occupation is studied in more detail, since it reportedly \cite{E17, D11} has a strong influence on the intrinsic damping. The possible coordination in the case of an ideal inverse spinel crystal structure, as it is given for undoped bulk \nife, is shown in Fig.\,\ref{figure5}(a). Half the Fe$^{3+}$ occupies octahedral, the other half tetrahedral sites, whereas Ni$^{2+}$ only sits at octahedral sites. Ideally, the dopant Zn$^{2+}$ would substitute for Ni$^{2+}$ and the smaller Al$^{3+}$ goes to octahedral Fe$^{3+}$ sites according to the nominal composition of Ni$_\text{0.65}$Zn$_\text{0.35}$Al$_\text{0.8}$Fe$_\text{1.2}$O$_\text{4}$ maintaining the inverse spinel. However, due to thin film growth on a normal spinel substrate MgAl$_2$O$_4$ a mixed spinel crystal structure leading to a small degree of A/B disorder, is most likely.\\
For experimental evidence, total electron yield XAS and XMCD spectra in \SI{20}{\degree} grazing incidence at the Ni and Fe L$_{3,2}$ edges shown in Fig.\,\ref{figure5} are recorded. Respective simulations are done with CTM4XAS \cite{SG10} using multiplet ligand field theory. The chosen parameters are adapted from nickel ferrite \cite{PLH11}, which shows similar XAS spectra and resulting XMCD curves. Scaling the percentages for the $F_{\text{dd}}$, $F_{\text{pd}}$, and $G_{\text{pd}}$, Slater integral reductions of \SI{70}{\percent} and \SI{80}{\percent}, respectively, are used to consider interatomic screening and mixing. For the octahedral coordination a crystal field splitting of $10Dq = $\,\SI{1.2}{\milli\electronvolt} and a positive exchange field of $J =  $\,\SI{48}{\milli\electronvolt} is used to fit the obtained Curie temperature of $T_{\text{C}} = (375 \pm 2)$\,\si{\kelvin} [see Fig.\,\ref{figure3}(c)] \cite{PMG09}. The tetrahedral coordination is simulated with a splitting of $10Dq = $\,\SI{-0.6}{\milli\electronvolt} and a negative exchange field $J = $\,\SI{-48}{\milli\electronvolt}. Charge transfer is not taken into account, because no influence on the absorption spectra has been found \cite{SG10}. Furthermore, the instrumental and intrinsic broadening is included by a Gaussian function of $\sigma = $\,\SI{0.25}{\electronvolt} and a Lorentzian function with a range of $\Gamma$[\SIrange{0.3}{0.5}{\electronvolt}], respectively. Note that the simulation is shifted in photon energy to fit the experimental spectra. \\
A comparison between the experimental and simulated XAS and XMCD spectra at the Ni L$_{3,2}$ edges [see Fig.\,\ref{figure5}(b)] suggests percentages of \SI{94}{\percent} Ni$^{2+}_{\text{Oh}}$, and \SI{4}{\percent} Ni$^{2+}_{\text{Td}}$, showing a slight deviation from the inverse spinel crystal structure. 
The assumption of A/B disorder is supported by evaluation of the simulation matching the experimental XAS and XMCD spectra of Fe [see Fig.\,\ref{figure5}(c)]. The Fe contributions are more complex, since not only Fe$^{3+}$ occupying Oh and Td sites but also typical spectroscopic signatures of Fe$^{2+}_{\text{Oh}}$ were identified in the experimental data. The best match between experiment and simulation was determined by relying on the main peaks of the XMCD at the L$_3$ edge as shown in Fig.\,\ref{figure5}(d). Percentages of \SI{33}{\percent} Fe$^{2+}_{\text{Oh}}$, \SI{45}{\percent} Fe$^{3+}_{\text{Oh}}$ and \SI{22}{\percent} Fe$^{3+}_{\text{Td}}$ are obtained as best fit. From this, a mixed state between the inverse and normal spinel crystal structure is apparent, as has been evidenced for thin film nickel ferrites in earlier works \cite{E17,K14,H15}. The element selective magnetic contributions from Ni and Fe indirectly infer the occupation of Zn and Al. According to the single ion model of ferrites \cite{D11}, Ni$^{2+}_{\text{Oh}}$ does not enhance the damping, but Ni$^{2+}_{\text{Td}}$ contributes by an unquenched orbital momentum. A non-negligible amount of \SI{4}{\percent} Ni$^{2+}_{\text{Td}}$ can be correlated with the reduced amount of Zn$^{2+}$ \cite{D11} as evidenced by EDX and RBS. Additionally, the strong imbalance between octahedral and tetrahedral coordinated Fe as well as the high amount of Fe$^{3+}_{\text{Oh}}$ suggest a deficiency of Al. This is further supported by the findings of bulk NiZAF reported by Ref.\,\cite{L13} since a higher amount of Al doping lowers the saturation magnetisation and the magnetic damping. 
However, the concentration of Al relative to Fe determined from RBS fits rather well with the nominal values. Therefore the difficulties of Fe incorporation cannot solely be explained or solved with the Al concentration. Additionally, the large amount of $\sim$ \SI{33}{\percent} Fe$^{2+}_{\text{Oh}}$ has a negative impact on the damping due to hopping [Fe$^{2+}$ $\rightarrow$ Fe$^{3+}$ + e\text{-}] \cite{E17,D11}. \\
The results are in good agreement with the previous analysis of the material system and give an explanation on the atomic level for the increased intrinsic damping. The main cause being a non-negligible amount of Ni$^{2+}_{\text{Td}}$ due to a low incorporation of Zn$^{2+}$ and large amounts of Fe$^{2+}_{\text{Oh}}$. The role of Ni$^{2+}_{\text{Td}}$ is even more important, since it directly influences the g-factor by orbital momentum. In addition, the strong A/B disorder promotes scattering centers resulting in the TMS contribution to magnetic damping.  Furthermore, the imbalance between tetrahedral and octahedral coordinated Fe$^{3+}$ supports the assumption of a mixed spinel structure as suggested by growth and chemical composition.

\section*{VI. Governing Mechanism}

\begin{figure*}[ht]
	\centering
	\includegraphics[width=0.84\textwidth]{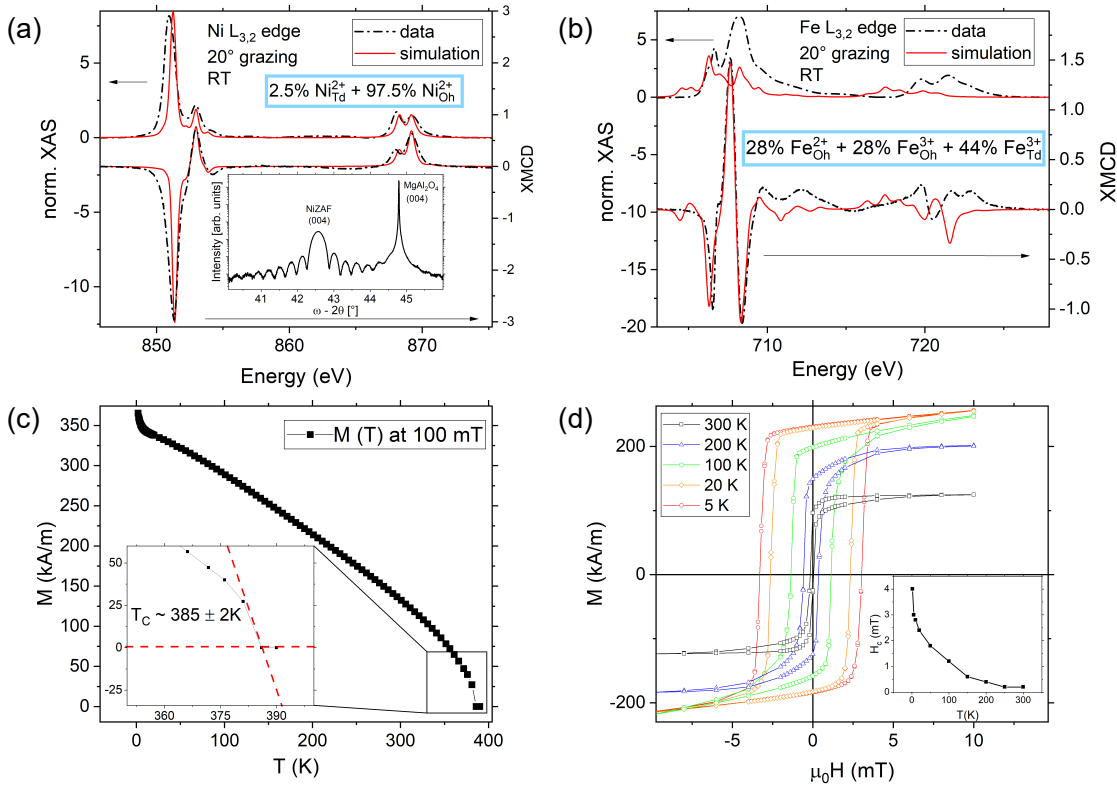}
	\vspace{-0.2cm}
	\caption{Analysis of a Zn-rich NiZAF showing the TEY XAS and XMCD spectra in grazing incidence of \SI{20}{\degree} at RT and simulations done with CTM4XAS \cite{SG10} at (a) the Ni L$_{3,2}$ and (b) the Fe L$_{3,2}$ edges. The inset in (a) depicts a symmetric $\omega - 2\theta$ scan. (c) the $M(T)$ curve at \SI{100}{mT} from \SIrange{2}{390}{\kelvin} with an inset magnifying the high temperature range. (d) shows hystereses at exemplary temperatures between \SI{300}{\kelvin} and \SI{2}{\kelvin} with an inset giving a detailed look at the coercive field in this temperature range.}
	\label{figure6}
\end{figure*} 

To evidence the crucial role of the Ni$^{2+}_{\text{Td}}$ in conjunction with the Zn deficiency a sample with a different composition of ($\text{Fe}:\text{Ni}:\text{Zn}:\text{Al}$)\,($1:0.43:0.39:0.71$) was fabricated. The composition was determined using RBS with a \SI{2}{\mega\electronvolt} He$^+$ beam in analogy to Fig.\,\ref{figure2}(c). For this purpose the Al content was kept constant to maintain strain and directly see the influence of the Zn dopant. By comparing the composition with the previous sample a strong change in the ($\text{Ni}:\text{Zn}$) ratio from ($0.62:0.26$) to ($0.43:0.39$) relative to Fe is apparent. Furthermore, the higher energetic \SI{10}{\mega\electronvolt} $^{12}$C$^+$ ion beam analysis of ($\text{Fe}:\text{Ni}:\text{Zn}:\text{Al}$) = ($1:0.48:0.42:0.76$) revealed an even more drastic change of the Ni-to-Zn ratio to ($0.48:0.42$). However, structural analysis shows a comparably excellent crystal quality with well pronounced Laue oscillations and a (within error bars similar or even slightly increased) perpendicular lattice parameter of $a_{\perp} = (8.49\pm 0.01)$\,\si{\angstrom} instead of  $a_{\perp} = (8.47\pm 0.01)$\,\si{\angstrom} [see inset of Fig.\,\ref{figure6}(a)]. Thus a tetragonal distortion of $c/a = 1.049 \pm 0.003$ is determined. \\
In Fig.\,\ref{figure6}(a) a comparison between data and simulation of the XAS and XMCD spectra at the Ni L$_{3,2}$ edges is depicted resulting in percentages of \SI{2.5}{\percent} Ni$^{2+}_{\text{Td}}$ and \SI{97.5}{\percent} Ni$^{2+}_{\text{Oh}}$. A reduction of almost \SI{50}{\percent} in tetrahedrally coordinated Ni either suggests an improved incorporation of Zn$^{2+}$ or an overabundance of Zn in this sample leading to less possibilities for Ni$^{2+}$ to occupy Td sites. Furthermore, the correlation between data and simulation of the XAS and XMCD spectra at the Fe L$_{3,2}$ edges shown in Fig.\,\ref{figure6}(b) confirms the results obtained for Ni. The simulation is fitted with respect to the primary peaks of the XMCD at the L$_3$ edge, since the magnetic information of the material is mainly contained in the XMCD spectra.  Percentages of \SI{28}{\percent} Fe$^{2+}_{\text{Oh}}$, \SI{28}{\percent} Fe$^{3+}_{\text{Oh}}$ and \SI{44}{\percent} Fe$^{3+}_{\text{Td}}$ are calculated indicating a more even distribution of Fe in the inverse spinel crystal structure in contrast to previous values [see Fig.\,\ref{figure5}(c)]. From this a better incorporation of the Al$^{3+}$ in the crystal structure can be inferred, even though the analysis of the chemical composition revealed the same amount of Al relatively to Fe. This further highlights the sensitivity of Fe to disorder in the crystal lattice. Additionally, the reduction of Fe$^{2+}_{\text{Oh}}$ suggests a positive impact on the intrinsic magnetic damping due to reduced hopping. This interpretation is confirmed by the improved magnetic properties of the material indicated by an increase in Curie temperature by \SI{10}{\kelvin} up to  $T_{\text{C}} = (385 \pm 2)$\,\si{\kelvin} at \SI{100}{\milli\tesla} as shown in the inset of Fig.\,\ref{figure6}(c).
Furthermore, the hystereses at exemplary temperatures in a range of \SIrange{300}{2}{\kelvin} shown in Fig.\,\ref{figure6}(d) indicate a transition to an even softer magnetic material than previously determined [see Fig.\,\ref{figure3}(a)]. From a closer look at the coercive fields dependence on the temperature depicted in the inset, a coercive field of $H_{\text{c}} \sim 0.2 $\,\si{\milli\tesla} at \SI{300}{\kelvin} is determined. At temperatures as low as \SI{100}{\kelvin} similar values for the coercivity as in previous samples of  $H_{\text{c}} = 1.2 $\,\si{\milli\tesla} are reached, increasing up to $H_{\text{c}} = 4 $\,\si{\milli\tesla} at \SI{2}{\kelvin}. Additionally, an increase of the saturation magnetization to  $M_{\text{S}} = (195 \pm 20 )$\,\si{\kilo\ampere/\meter} at RT  (not shown, see repository \cite{LBN19}) in comparison to $M_{\text{S}} = (118 \pm 12)$\,\si{\kilo\ampere/\meter} was measured.

\begin{figure}[h]
	\centering
	\includegraphics[width=0.41\textwidth]{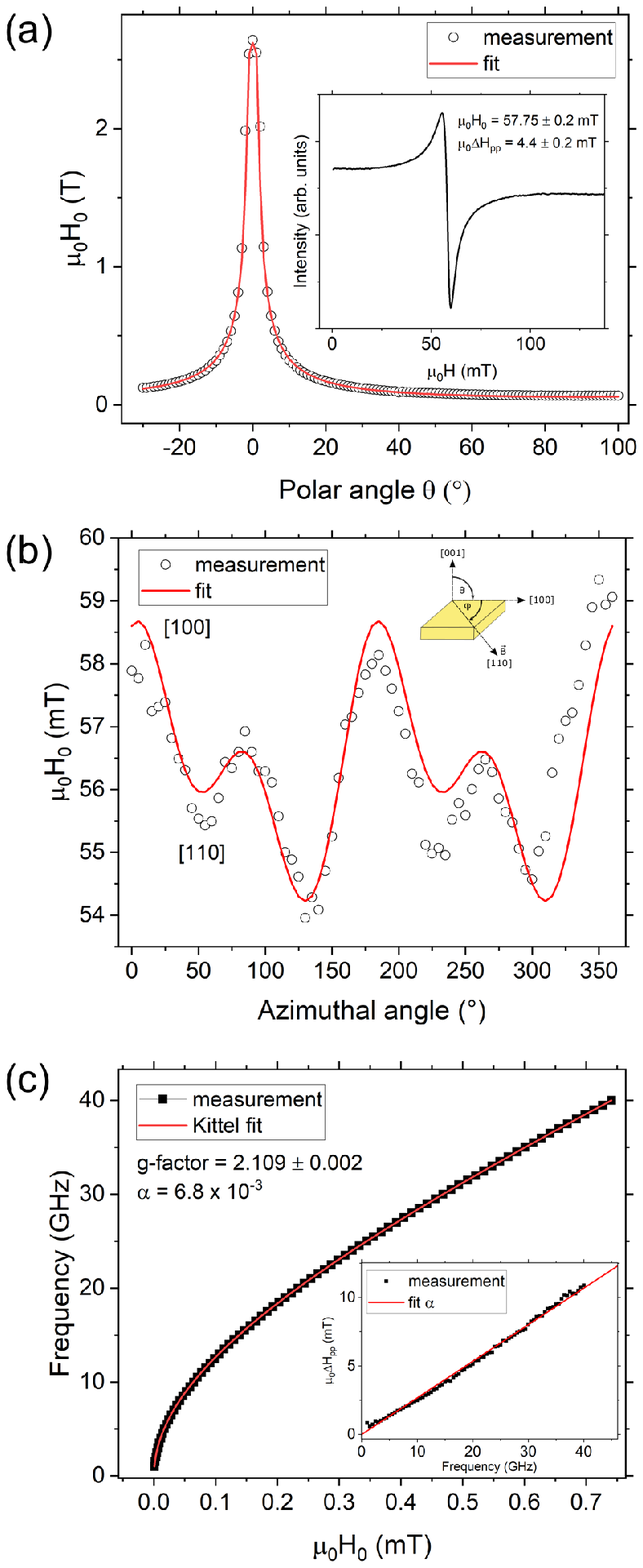}
	\vspace{-0.2cm}
	\caption{Dynamic magnetic properties including (a) the polar angular dependence from \SIrange{-20}{90}{\degree} and (b) the azimuthal angular dependence from \SIrange{0}{360}{\degree} with an anisotropy fit (red line) for the resonance position in both dependences. The inset in (a) depicts the smallest FMR line at $f= 9.5$\,\si{\giga\hertz} in the easy axis. (c) shows the frequency dependence of the resonance position as well as line-width (inset) in a range from \SIrange{5}{40}{\giga\hertz}.}
	\label{figure7}
\end{figure}

In Fig.\,\ref{figure7} the VNA-FMR analysis is shown. In the inset of Fig.\,\ref{figure7}(a) the FMR line at $f = $\,\SI{9.5}{\giga\hertz} is shown indicating an increased resonance position of $\mu_0 H = (57.8 \pm 0.2)$\,\si{\milli\tesla} and a reduced linewidth of $\mu_0 \Delta H_{\text{pp}} = (4.4 \pm 0.2)$\,\si{\milli\tesla} confirming the improved magnetic properties. Since the resonance position depends on the magnetocrystalline anisotropy this coincides with the decreased anisotropy of $2K_{2\perp}/M_\mathrm{s} = 1.53$\,\si{\tesla} determined by the polar angular dependence. Additionally, a lower effective magnetization of $\mu_0 M_{\text{eff}} = $\,\SI{1.78}{\tesla} is calculated from the fit to the angular dependences.
The azimuthal angular dependence confirms the cubic (fourfold) symmetry of the crystal. The easy axis in the $\left\langle 110\right\rangle$ direction coincides with NiZAF [see Fig.\,\ref{figure4}(b)], however the cubic in plane anisotropy has two components $2K_{4\perp}/M_\mathrm{s} = - (420 \pm 100)$\,\si{\milli\tesla} and $2K_{4\parallel}/M_\mathrm{s} = - 1.2$\,\si{\milli\tesla}, indicating a slight shift towards a tetragonal crystal symmetry. Additionally, a slight uniaxial in plane anisotropy component of $2K_{\parallel}/M_\mathrm{s} = - 1.4$\,\si{\milli\tesla} is needed to fit the data.
The frequency dependences of the resonance position and the linewidth are shown in Fig.\,\ref{figure7}(c) and its inset, respectively. Due to the larger FMR signal, the g-factor could be determined with high precision yielding $g = 2.109 \pm 0.002$. A comparison with previous work \cite{E17}, where values of $g = 2.29 \pm 0.09$ are obtained, shows a drastic change in orbital momentum. These findings match the decreased Ni$^{2+}_{\text{Td}}$, since these contribute to damping by unquenched orbital momentum. Furthermore, the frequency dependence of the linewidth shows that in the Zn-rich sample no two magnon scattering is apparent, resulting in a Gilbert damping of $\alpha = 6.8 \times 10^{-3}$. Additionally, the lack of an inhomogeneous contribution suggests an excellent crystal growth without any defects.\\
Although the strain has slightly increased, by improving the cation distribution and deviating from the reported best stoichiometry \cite{E17,L13} the magnetic anisotropy and the linewidth broadening effects could be reduced remarkably. These results demonstrate the relevance of the cation distribution, in particular the amount of Ni$^{2+}_{\text{Td}}$ and Fe$^{2+}_{\text{Oh}}$ as another major criteria for the optimisation of the magnetic properties. In particular, the reduction of Ni$^{2+}_{\text{Td}}$ and  A/B disorder, lowered the g-factor and magnetic damping and eliminated the TMS contribution.\\
So far, only the influence of Al on bulk NiZAF \cite{L13} has been studied systematically. The present investigation shows the importance of tuning both the Al and the Zn content. Therefore, for a full understanding of this complex material system the determination of the whole parameter space of Al and Zn concentration together with the site occupancy in NiZAF thin films is required.
\section*{VII. Conclusion}

In this work the growth of Zn/Al doped nickel ferrite thin films with excellent crystal quality was achieved by reactive magnetron sputtering from a target with nominal composition of Ni$_\text{0.65}$Zn$_\text{0.35}$Al$_\text{0.8}$Fe$_\text{1.2}$O$_\text{4}$. The chemical composition of ($\text{Fe}:\text{Ni}:\text{Zn}:\text{Al}$)\,($1:0.62:0.26:0.71$) obtained by RBS evidenced a rather good stoichiometric transfer with a slight Zn deficiency. The material is highly, but coherently strained and no indication for defects or dislocations can be observed from a structural analysis. A comparison with bulk \cite{L13} shows a shift of the in-plane magnetic easy axis from $\left\langle 111\right\rangle$ to the $\left\langle 110\right\rangle$ in agreement with the thin films in \cite{E17}. Furthermore, the frequency dependence revealed a reduced g-factor, an increased damping and additional contributions from two magnon scattering and inhomogeneity compared to Ref.\,\cite{E17} indicating a lower amount of orbital momentum.\\
These changes are caused by the cation distribution, which has a significant influence on the magnetocrystalline anisotropy and the intrinsic magnetic damping. A comparative sample with increased Zn content shows a reduction of Ni$^{2+}_{\text{Td}}$, which infers a better incorporation of Zn$^{2+}$ leading to a decrease in coercivity, anisotropy and damping. Due to the reduction in orbital momentum the g-factor is lowered significantly. Furthermore, less Fe$^{2+}_{\text{Oh}}$ resulted in a smaller linewidth and an even lower magnetic damping. By improving the A/B disorder, the contributions of two magnon scattering and inhomogeneity could be eliminated.\\
To conclude, in contrast to initial assumptions, strain is not the sole mechanism to control the magnetic properties in this complex material system. In addition, the cation distribution, i. e. the amount of Fe$^{2+}_{\text{Oh}}$ and Ni$^{2+}_{\text{Td}}$ was found to have a major impact on the magnetic anisotropy and damping independent of strain. 
The influence of the doping can be divided into two aspects: To first order the Ni:Zn ratio controls the magnetic damping and g-factor and the Fe\,:\,Al ratio the strain. Secondly, the resulting A/B disorder promotes the formation of scattering centers leading to TMS contributions and inhomogeneity.
While previous work on bulk NiZAF have already evidenced the importance of the Al concentration, the present work demonstrates that for future optimization the crucial role of the Zn concentration has to be taken into account.

\section*{Acknowledgement}

The authors gratefully acknowledge funding by FWF project ORD-49. The X-ray absorption measurements were performed on the EPFL/PSI X-Treme beamline at the Swiss Light Source, Paul Scherrer Institut, Villigen, Switzerland. In addition, support by VR-RFI (Contract No. 2017-00646\_9) and the Swedish Foundation for Strategic Research (SSF, Contract No. RIF14-0053) supporting accelerator operation at Uppsala University is gratefully acknowledged. The financial support by the Austrian Federal Ministry for Digital and Economic Affairs and the National Foundation for Research, Technology and Development in the frame of the CDL for Nanoscale Phase Transformations is gratefully acknowledged. Further the authors thank Werner Ginzinger for TEM sample preparation.

\end{document}